\documentclass{ws-p8-50x6-00}
\usepackage{graphicx,wrapfig}

\begin{document}

\title{Measurements of Higher Order Effects in QCD from the Tevatron
  Collider}

\author{Thomas Nunnemann\\
For the D\O\ and CDF Collaborations}

\address{Fermi National Accelerator Laboratory, P.O.~Box 500,
Batavia, IL 60510, U.S.A.\\ 
E-mail: nunne@fnal.gov}

\maketitle

\abstracts{Quantum Chromodynamics has been studied extensively at 
  Fermilab's Tevatron collider. Between 1992 and 1996 the D\O\ and
  CDF experiments, each accumulated approximately 
  $100\,\mathrm{pb}^{-1}$ of proton-antiproton collisions at
  a center-of-mass energy $\sqrt{s}=1800\,\mathrm{GeV}$ and 
  $\sim 0.5\,\mathrm{pb}^{-1}$ at $\sqrt{s}=630\,\mathrm{GeV}$. In this
  paper, we present selected recent measurements of higher order effects in
  QCD: multiple jet production and subjet and charged particle multiplicities
  in quark and gluon jets.}

\section{Introduction}
Within the framework of Quantum Chromodynamics (QCD) the inelastic scattering
between a proton and an antiproton can be described as a hard interaction
between their partons, namely quarks and gluons. After the collision, the
outgoing partons fragment and hadronize in streams of particles called jets.
Whereas lowest order perturbative QCD only predicts the production of
two partons, which are emitted at opposite transverse momentum, higher order
effects account for the production of multiple jets and the internal structure
of jets.
In this paper, we summarize recent measurements by the D\O\ and CDF
collaborations based on approximately 
$100\,\mathrm{pb}^{-1}$ of proton-antiproton collision data at
a center-of-mass energy $\sqrt{s}=1800\,\mathrm{GeV}$ and 
$\sim 0.5\,\mathrm{pb}^{-1}$ at $\sqrt{s}=630\,\mathrm{GeV}$ obtained 
between 1992 and 1996: ratios of inclusive multijet cross 
sections,\cite{D0:R32} 
as well as subjet
and charged particle multiplicities in quark and gluon initiated 
jets.\cite{D0:subjet,CDF:multi}

\subsection{Jet Reconstruction}
Jets are commonly defined by their energy deposition in multiple calorimeter
towers.
For the reconstruction of jets, most of the analyses presented here use an
iterative fixed cone algorithm with a cone radius of typically
$\mathcal{R}=0.7$ in $\eta-\phi$ space (pseudo-rapidity is defined as
$\eta = \ln (\tan\frac{\theta}{2}$)).\cite{jetprd}
The D\O\ subjet multiplicity
measurement applies a successive combination algorithm
based on relative transverse momenta, the $k_T$ algorithm
with a clustering parameter
$\mathcal{D}=0.5$.\cite{D0:subjet} 
CDF exploits tracks reconstructed within fixed cone sizes
for their analysis of charged particle multiplicities.

\section{Ratios of Multijet Cross Section}
Three-jet production probes the rate of gluon emission, thus the fundamental
coupling of QCD: $\alpha_s$. In addition its study allows to investigate
possible scale differences for the secondary gluon vertex as compared to the
initial hard interaction.

D\O\ measures the ratio of inclusive three-jet to two-jet production
$$
R_{32} = \frac{\sigma_3}{\sigma_2} =
\frac{\sigma(p\bar{p} \rightarrow \ge 3 jets + X)}{\sigma(p\bar{p}
\rightarrow \ge 2 jets + X)}
$$
as a function of the scalar sum of transverse energy
$H_T=\sum E_T^{\mathrm{jet}}$.

\begin{wrapfigure}{r}{0.5\textwidth}
  \centering
  \includegraphics[width=0.5\textwidth]{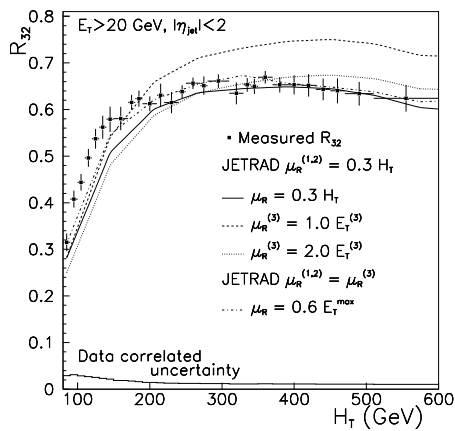}
  \caption{D\O\ measurement of the ratio of inclusive three-jet to inclusive 
    two-jet cross section $R_{32}$ as a function of the scalar sum of jet 
    transverse energies $H_T$. The four smoothed distributions show the
    {\sc Jetrad} prediction for the renormalization scales
    indicated in the legend.
    \label{fig:R32}}
\end{wrapfigure}

The measurement of $R_{32}$ as a function of $H_T$ is performed for different
jet $E_T$ thresholds of 20, 30, and 40\,GeV
and within the ranges $|\eta_{\mathrm{jet}}|<3$ and 
$|\eta_{\mathrm{jet}}|<2$.\cite{D0:R32}
The results for jet $E_T>20\,\mathrm{GeV}$ and 
central pseudo-rapidities
are shown in Figure~\ref{fig:R32}.
To study the sensitivity to the choice of renormalization scale $\mu_R$ for
the production of additional jets,
the data are compared to next-to-leading order
(NLO) QCD predictions of {\sc Jetrad}\cite{jetrad} for different assumptions
for the scales $\mu_R^{(i)}$ of the two vertices.
For all {\sc Jetrad} predictions the factorization scale $\mu_F$ has been
chosen to be identical to $\mu_R$.
The scale $\mu_R^{(3)}$ for the third jet has been varied proportionally to 
$H_T$, the transverse energy of the third
jet $E_T^{(3)}$ and the maximum jet transverse energy $E_T^{\mathrm{max}}$,
while the scale for the two leading jets $\mu_R^{(1,2)}$ has been set either
proportional to $H_T$ or $E_T^{\mathrm{max}}$.
A fair description of the data is obtained for the choices 
$\mu_R^{(1,2,3)} \sim 0.3\, H_T$ and
$\mu_R^{(1,2,3)} \sim 0.6\, E_T^{\mathrm{max}}$.
The result does not indicate a need for a significantly different scale
for the 
secondary gluon vertex from that of the primary hard interaction.



\section{Subjet and Charged Particle Multiplicities in Quark and Gluon Jets}
Asymptotically, the ratio of the number of particles within gluon 
initiated jets to
that in quark jets is expected to be in the ratio of their color charges
$C_A/C_F = 9/4$. Corrections to this estimate are given by
perturbative QCD calculations within the framework of the
Modified Leading Log Approximation.\cite{CDF:multi} 
Measurements of charged particle and subjet multiplicities not only allow 
to test the applicability of perturbative QCD calculations to the description
of the soft process of jet fragmentation, but might also provide a method
of jet tagging in other analyses.

Complementary to subjet multiplicity measurements at LEP\cite{subjetslep} and 
HERA,\cite{subjetshera}
D\O\ and CDF are using two different techniques to select
quark and gluon enriched samples based on the average momentum fraction $x$
of the initial partons (low $x$: gluon dominance, high $x$: valence quarks).
The relative quark and gluon content in the enriched samples is inferred from
the {\sc Herwig} Monte Carlo generator\cite{herwig} and the 
CTEQ4M (D\O\ and CDF) and CTEQ4HJ (CDF)
parton distribution functions.

\subsection{Subjet Multiplicity at D\O}
\begin{figure}
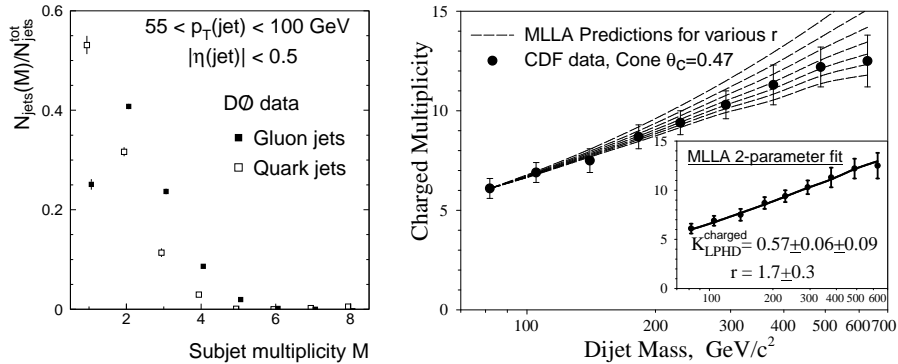

  \centering
  \mbox{\parbox{4.8cm}{\includegraphics[width=4.8cm]{psfiles/pub4b_29.epsi}}
  \hspace{0.3cm}
  \parbox{6.5cm}{\includegraphics[width=6.45cm]{psfiles/pub5_1.epsi}}}
  \caption{Left: Corrected (particle-level) subjet multiplicity for gluon and
    quark jets, extracted from D\O\ inclusive jet data at 
    $\sqrt{s} = 1800\,\mathrm{GeV}$
    and $630\,\mathrm{GeV}$. 
    Jets and subjets are defined with the $k_T$ algorithm.
    Right: Average multiplicity of charged particles per jet vs. dijet mass.
    The curves show MLLA prediction for different values of $1\le r \le 2.25$
    (increasing from top to bottom). The two-parameter MLLA fit is
    represented by the solid line in the insert.
    \label{fig:multiplicities}}
\end{figure}

D\O\ measures the subjet multiplicity for center-of-mass energies 
$\sqrt{s}=1800$\,GeV and 630\,GeV at fixed $E_T$ and $\eta$,\cite{D0:subjet}
effectively
probing smaller values of $x$ with increasing $\sqrt{s}$.
Jets and subjets are defined using the $k_T$ algorithm with a clustering
parameter $\mathcal{D}=0.5$ and a resolution parameter
$y_{\mathrm{cut}}=10^{-3}$.
The unfolded subjet multiplicities for quark and gluon jets, 
corrected to particle
level, are shown in the left panel of Figure~\ref{fig:multiplicities}.
The ratio of the average multiplicities $M_{q/g}$ 
(reduced by one to account for the initial parton) is measured to be
$R={(<M_g>-1)}/{(<M_q>-1)} = 
1.84 \pm 0.15\,\mathrm{(stat.)}\,^{+0.22}_{-0.18}\,\mathrm{(sys.)}$, 
in agreement with the {\sc Herwig} prediction of $R=1.91$.

\subsection{Charged Particle Multiplicities at CDF}
CDF measures the mean charged particle multiplicity within fixed cones of 
$\theta_c=$ $0.17$, $0.28$ and $0.47$ in dijet events as a function of dijet
mass between $80$ and $630\,\mathrm{GeV}/c^2$,\cite{CDF:multi}
which is proportional to the product of the momentum contributions of the
incoming partons.
A fit to the data (cf.~Fig.~\ref{fig:multiplicities}, right) 
within the framework of the 
Modified Leading Log Approximation (MLLA)\cite{mlla}
and assuming 
Local Parton Hadron Duality (LPHD)\cite{lphd}
yields
$r={<M_g>}/{<M_q>} = 1.7 \pm 0.3$ 
for the ratio of parton multiplicities,
in agreement with perturbative QCD calculations,
and for the ratio of multiplicities of charged hadrons to partons
$K_\mathrm{LPHD}^\mathrm{charged}=
{N^\mathrm{charged}_\mathrm{hadrons}}/{N_\mathrm{partons}}=
0.57 \pm 0.11$\,.
After normalization ($\times 0.89$) {\sc Herwig} gives a good description
of the evolution of charged multiplicity as function of dijet mass.

\end{document}